\documentclass[twocolumn,showpacs,prb,amsmath,amssymb]{revtex4}
\usepackage{graphicx}
\usepackage{dcolumn}
\usepackage{bm}

\begin{document}

\title{High-order density-matrix perturbation theory}
\author{Michele Lazzeri and Francesco Mauri}
\affiliation{Laboratoire de Min\'eralogie Cristallographie de Paris,
Paris, France.}
\date{\today}

\begin{abstract}
We present a simple formalism for the calculation of the derivatives of the
electronic density matrix $\rho$ at any order, within density
functional theory.
Our approach, contrary to previous ones, is not based on the
perturbative expansion of the Kohn-Sham wavefunctions.
It has the following advantages:
(i) it allows a simple derivation for the expression for the high order
derivatives of $\rho$;
(ii) in extended insulators, the treatment of uniform-electric-field
perturbations and of the polarization derivatives is straightforward.
\end{abstract}

\pacs{ 71.15.-m,71.15.Mb}
\maketitle

\section{Introduction}
Linear response methods,~\cite{DFPT1,DFPT2} within the density functional
theory approach (DFT),~\cite{DFT} have been successfully
applied to compute a wide range of properties
in real materials such as phonon dispersions, dielectric constants,
effective charges,~\cite{DFPT1,DFPT2} and NMR spectra.~\cite{NMR}

Beyond linear response, perturbation theory applied to the Kohn-Sham
(KS) orbitals allows the calculation of the derivatives of the energy at
any order.~\cite{gonze95}
This kind of approach has two disadvantages:
(i) although the final result is
gauge invariant, {\it i.e.} invariant with respect to an arbitrary
unitary rotation in the space of the occupied KS-orbitals,~\cite{gonze95}
the formulation
of the theory depends on the chosen gauge. This becomes apparent in the
application of the KS-orbitals orthonormality constraints at high order.
(ii) In the case of periodic systems, the treatment of a perturbation due
to a uniform electric field is not trivial, because the position operator,
necessary to describe such a perturbation, is ill-defined in
periodic boundary conditions.
Much effort has been devoted throughout the years to overcome this last
problem. Early treatments of the electric field perturbation
for the calculation of the second and third order
susceptibilities are particularly complex.~\cite{levine}
A simpler formalism for the calculation of the second order susceptibility was
obtained in Ref.~\onlinecite{dalcorso}, taking advantage of the $2n+1$ theorem
and of a Wannier representation of the orbitals.
Only very recently Nu$\tilde{\rm n}$es and Gonze~\cite{nunes01} were able to
give an expression for the derivatives of the DFT energy, with
respect to uniform electric fields, at any order, by introducing in the
Hamiltonian an
additional term depending on the polarization Berry phase.~\cite{kingsmith93}

We remark that, although a perturbation due to a macroscopic uniform electric
field is ill-defined on an individual Bloch states, such a perturbation is well
defined on individual Wannier states,~\cite{wannier,dalcorso}
which can be obtained by a different choice of gauge.
This consideration suggest that the two problems, mentioned in the previous
paragraph, are related, and that both problems might possibly disappear using
a perturbative approach which is {\it not} based on the perturbative series
of the single KS orbitals, but is solely based on the properties of the
electronic density matrix $\rho$, which is a gauge independent operator.

In a recent paper~\cite{lazzeri03} we gave an expression for the second
order derivative of $\rho$ which allowed
the efficient computation of Raman spectra.~\cite{lazzeri03}
In the present paper we derive a general expression for the n-th order
derivative of $\rho$, using the two relations $\rho^2=\rho$, and
$[\rho,H]=0$, being $[,]$ a commutator and $H$ the KS Hamiltonian.

To fix our notation we define the electronic density matrix as
\[
\rho=\sum_v |\psi_v\rangle\langle\psi_v|,
\]
where, throughout the paper, $v$ or $v'$ is an index running
on the occupied valence states, and $|\psi_v\rangle$ are normalized
KS-eigenstates,
{\it i.e.} $H |\psi_v\rangle=\epsilon_v|\psi_{v}\rangle$.
Given a perturbation associated with a small parameter $\lambda$,
for a generic quantity $F$, we consider the perturbation series:
\begin{equation}
F(\lambda) = F^{(0)} + \lambda F^{(1)} + \lambda^2 F^{(2)}
+ \lambda^3 F^{(3)} + \dots.
\label{series}
\end{equation}
The generalization to the case of different perturbations
$\lambda_1, \dots \lambda_n$ is straightforward.
$\rho$ and $H$ stand for $\rho(\lambda)$ and $H(\lambda)$.
We call $P_V$, and $P_C$, respectively, the projectors on valence and
conduction band states, $i.e.~P_V=\rho^{(0)}, P_C={\bf 1}-\rho^{(0)}$.
Given an Hermitean operator $A$ we define
$A_{CC}=P_CAP_C,~A_{VV}=P_VAP_V,A_{CV}=P_CAP_V$, and
$A_{VC}=(A_{CV})^\dagger$.

The work is organized as follows.
In Sec.~\ref{sec1}, we use the relation $\rho^2=\rho$ to express $\rho^{(n)}$
as a function of the operators $\rho^{(i)}_{CV}$, having $i\leq n$.
In Sec.~\ref{sec2}, we use the relation $[H,\rho]=0$
to obtain an expression for $\rho^{(n)}_{CV}$ that can be easily computed
using standard linear response techniques.
In Sec.~\ref{sec3}, we show that, within our formalism, the perturbations
due to a uniform electric field are well defined in extended insulators.
In Sec.~\ref{sec4}, we derive a simple expression for the
derivatives of the polarization.

\section{$\rho^{(n)}$ as a function of $\{\rho^{(i)}_{CV}\}$, with $i\leq n$
\label{sec1}}
We decompose $\rho$ in $\rho_{CC}+\rho_{VV}+\rho_{CV}+\rho_{VC}$, and
we consider these four terms separately.
The idempotency condition, $\rho^2=\rho$, implies that
$P_C\rho P_C=P_C\rho\rho P_C=P_C\rho (P_C+P_V)\rho P_C$,
or
\[
\rho_{CC}-\rho_{CC}\rho_{CC} = \rho_{CV}\rho_{VC}.
\]
When all the eigenvalues of $\rho_{CC}$ are lower than $1/2$,
{\it i.e.} for $\lambda$ sufficiently small, 
this relation between the two operators $\rho_{CC}$ and
$\rho_{CV}\rho_{VC}$ can be inverted to obtain:
\begin{equation}
\rho_{CC}=\frac{1-\sqrt{1-4\rho_{CV}\rho_{VC}}}{2},
\label{rhocc}
\end{equation}
where the right-hand side denotes the operator obtained substituting
$\rho_{CV}\rho_{VC}$ in the Taylor series
\begin{equation}
\frac{1-\sqrt{1-4x}}{2} = x +x^2 +2x^3 +5x^4 +\dots.
\label{taylor}
\end{equation}
In a similar way, defining $\Delta\rho_{VV}=\rho_{VV}-\rho^{(0)}$,
\[
\Delta\rho_{VV}\Delta\rho_{VV}+\Delta\rho_{VV}=
\rho_{VV}\rho_{VV}-\rho_{VV}=-\rho_{VC}\rho_{CV}.
\]
When all the eigenvalues of $\Delta\rho_{VV}$ are larger than $-1/2$,
{\it i.e.} for $\lambda$ sufficiently small,
$\Delta\rho_{VV}$ can be expressed as a function of $\rho_{VC}\rho_{CV}$:
\begin{equation}
\Delta\rho_{VV}=\rho_{VV}-\rho^{(0)}
=-\frac{1-\sqrt{1-4\rho_{VC}\rho_{CV}}}{2}.
\label{rhovv}
\end{equation}

Finally, $\rho^{(n)}$ can be expressed as a function of the
$\{\rho^{(i)}_{CV}\}$, with $i\leq n$, using the relation
\begin{equation}
\rho^{(n)}=\rho^{(n)}_{CV}+
\rho^{(n)}_{VC}+\rho^{(n)}_{CC}+\rho^{(n)}_{VV},
\label{rhondec}
\end{equation}
and taking the $n$-th order variation of Eq.~(\ref{rhocc})
and Eq.~(\ref{rhovv}) through Eq.~(\ref{taylor}).
As examples, observing that $\rho^{(0)}_{CV}=0$,
is easy to show that
\begin{eqnarray}
\label{eqr1}
\rho^{(1)}&=&\rho^{(1)}_{CV}+\rho^{(1)}_{VC} \\
\label{eqr2}
\rho^{(2)}&=&\rho^{(2)}_{CV}+\rho^{(2)}_{VC}+
[\rho^{(1)}_{CV},\rho^{(1)}_{VC}] \\
\label{eqr3}
\rho^{(3)}&=&\rho^{(3)}_{CV}+\rho^{(3)}_{VC}+
[\rho^{(1)}_{CV},\rho^{(2)}_{VC}]+
[\rho^{(2)}_{CV},\rho^{(1)}_{VC}] \\
\label{eqr4}
\rho^{(4)}&=&\rho^{(4)}_{CV}+\rho^{(4)}_{VC}+
[\rho^{(1)}_{CV},\rho^{(3)}_{VC}]+[\rho^{(2)}_{CV},\rho^{(2)}_{VC}]+
\nonumber \\
&&+[\rho^{(3)}_{CV},\rho^{(1)}_{VC}]
+[\rho^{(1)}_{CV}\rho^{(1)}_{VC}\rho^{(1)}_{CV},\rho^{(1)}_{VC}].
\end{eqnarray}
Note that each $\rho^{(i)}_{CV}$ is a gauge independent operator.

In Eqs.~(\ref{eqr1})-(\ref{eqr4}), $\rho^{(n)}$ is expressed as
$\rho^{(n)}_{CV}+\rho^{(n)}_{VC}$ plus a commutator, for $n\leq 4$.
This property is used in Sec.~\ref{sec4}
to compute the derivatives of the polarization.
It holds at any order $n$. Indeed, as we show in the appendix:
\begin{eqnarray}
\rho^{(n)}&=&\rho^{(n)}_{CV}+\rho^{(n)}_{VC}+
\sum_{i=1}^{n-1}[\rho^{(i)}_{CV},O^{(n-i)}_{VC}] = \nonumber\\
&=&\rho^{(n)}_{CV}+\rho^{(n)}_{VC}+
\sum_{i=1}^{n-1}[O^{(n-i)}_{CV},\rho^{(i)}_{VC}],
\label{rhon}
\end{eqnarray}
where $n\geq 2$, and
\begin{eqnarray}
O_{VC}&=&\rho_{VC}\frac{1-\sqrt{1-4\rho_{CV}\rho_{VC}}}{2\rho_{CV}\rho_{VC}}=
\nonumber \\
&=&\frac{1-\sqrt{1-4\rho_{VC}\rho_{CV}}}{2\rho_{VC}\rho_{CV}}\rho_{CV}.
\label{ocv}
\end{eqnarray}

\section{Computation of $\rho^{(n)}_{CV}$
\label{sec2}}
In order to compute $\rho^{(n)}_{CV}$ we introduce the wavefunction
$|\eta^{(n)}_v\rangle=P_C\rho^{(n)}|\psi^{(0)}_v\rangle$,
$|\psi^{(0)}_v\rangle$ being an unperturbed KS eigenvector.
We have:
\begin{equation}
\rho^{(n)}_{CV}=\sum_vP_C\rho^{(n)}|\psi^{(0)}_v\rangle\langle\psi^{(0)}_v
|=\\
\sum_v |\eta^{(n)}_v\rangle\langle\psi^{(0)}_v|.
\label{eq4}
\end{equation}
Equating to zero the n-th order term of the perturbation
series of $[H,\rho]=0$, we find:
\[
\sum_{i=0}^{n} [H^{(i)},\rho^{(n-i)}]=0.
\]
Multiplying this relation on the left by $P_C$ and applying to
$|\psi^{(0)}_v\rangle$ to the right, we derive:
\begin{equation}
\label{lin_sys}
\left(H^{(0)}-\epsilon^{(0)}_v\right) 
|\eta^{(n)}_v\rangle=
-\sum_{i=1}^{n}P_C[H^{(i)},\rho^{(n-i)}]|\psi^{(0)}_v\rangle.
\end{equation}
Solving the linear system of Eq.~(\ref{lin_sys}) one can obtain
$|\eta^{(n)}_v\rangle$ and, thus, $\rho^{(n)}$.
Since the right-hand side of Eq.~(\ref{lin_sys})
depends on $H^{(n)}$, that in turn depends on $\rho^{(n)}$, the system
is to be solved self-consistently, {\it e.g.} by using an iterative
procedure.
Eq.~(\ref{lin_sys}) needs to be solved only
for a finite number of $|\eta^{(n)}_v\rangle$ functions, running the index $v$
on the sole valence states.
The linear system of Eq.~(\ref{lin_sys}) is analogous to the one that is to be
solved in the standard density functional perturbation theory (DFPT),
~\cite{DFPT1,DFPT2}
thus Eqs.~(\ref{rhon},\ref{eq4},\ref{lin_sys}) give an efficient algorithm
that can be easily implemented in available DFPT codes (as the
PWSCF~\cite{pwscf} or the ABINIT~\cite{abinit} code),
to compute the derivatives of $\rho$ at any order.

Alternatively, Eq.~(\ref{lin_sys}) can be written as
\begin{equation}
|\eta^{(n)}_v\rangle = \tilde G_v
\left(\sum_{i=1}^{n}[H^{(i)},\rho^{(n-i)}]\right)|\psi^{(0)}_v\rangle,
\label{eq6}
\end{equation}
where
$\tilde G_v = \sum_c |\psi^{(0)}_c\rangle\langle\psi^{(0)}_c|/
(\epsilon^{(0)}_v-\epsilon^{(0)}_c)$
is the unperturbed Green function operator projected on the conduction band,
and the sum $\sum_c$ is restricted to the empty conduction-band states.
From Eq.~(\ref{eq6}) one can recognize that
$|\eta^{(1)}_v\rangle=P_C|\psi^{(1)}_v\rangle$ and that
$|\eta^{(2)}_v\rangle$ is the projection on the conduction band of the KS
orbital $v$ in the parallel-transport
gauge of Ref.~\onlinecite{gonze95}.
However, at higher orders, there is not such a simple relation between 
the $|\eta^{(i)}_v\rangle$ functions, defined in the present paper, and
the variations of the KS-orbitals.

Finally, as examples, we write $\rho^{(i)}$ as a function of the
$|\eta^{(i)}_v\rangle$ wavefunctions, for the three lowest order:
\begin{eqnarray}
\rho^{(1)}&=&\sum_v\left(
|\eta^{(1)}_v\rangle\langle\psi^{(0)}_v| + 
|\psi^{(0)}_v\rangle\langle\eta^{(1)}_v| \right)\\
\rho^{(2)}&=&\sum_v \left(|\eta^{(2)}_v\rangle\langle\psi^{(0)}_v| +
|\psi^{(0)}_v\rangle\langle\eta^{(2)}_v|+
|\eta^{(1)}_v\rangle\langle\eta^{(1)}_v| \right) +\nonumber \\
&&
-\sum_{v,v'}|\psi^{(0)}_v\rangle\langle\eta^{(1)}_v|\eta^{(1)}_{v'}\rangle
\langle\psi^{(0)}_{v'}|\phantom{}\label{rhopsi}\\
\rho^{(3)}&=&\left(
\sum_v|\eta^{(3)}_v\rangle\langle\psi^{(0)}_v| +
\sum_v|\eta^{(2)}_v\rangle\langle\eta^{(1)}_v| +\right.\nonumber \\
&& - \left. \sum_{v,v'}
|\psi^{(0)}_v\rangle\langle\eta^{(2)}_v|\eta^{(1)}_{v'}\rangle
\langle\psi^{(0)}_{v'}|\right) + (\dots)^\dagger.
\end{eqnarray}
We already used Eq.~(\ref{rhopsi}) in Ref.~\onlinecite{lazzeri03}, to
compute the Raman tensor.

\section{Treatment of the electric fields
\label{sec3}}
Thanks to the commutators in Eq.~(\ref{lin_sys}), all the quantities needed to
compute $\rho^{(n)}$ are well defined in an extended insulator, even if the
perturbation $\lambda$ is the component $E_\alpha$ of a uniform electric field,
{\it i.e.}, if
$H^{(1)} = -e r_\alpha + \partial V^{\rm Hxc}/\partial{E_\alpha}$,
~\cite{nota2} being
$r_\alpha$ the $\alpha^{th}$ Cartesian component of the position operator
${\bf r}$, $e$ the electron charge,
and $V^{\rm Hxc}$ the Hatree and exchange-correlation potential.
In particular, in an insulator, the
commutator $[{\bf r},\rho^{(n-1)}]$, which appears in
Eq.~(\ref{lin_sys}), is a well-defined  bounded  operator, since the 
variation of the density matrix
is localized ($\langle {\bf r}'' |\rho^{(n-1)}| {\bf r}' \rangle$ goes
to zero exponentially for $|{\bf r}''-{\bf r}'|\rightarrow \infty$).

To prove the localized nature of $\rho^{(n-1)}$ in a periodic system,
we notice that $\rho^{(n-1)}$ can be written (see Eq.~(\ref{rhopsi})) as
a sum of operators of the type
\[
D=\sum_{{\bf k}v}
|\alpha_{{\bf k}v}\rangle\langle\beta_{{\bf k}v}|,
\]
where
$|\alpha_{{\bf k}v}\rangle$ and $|\beta_{{\bf k}v}\rangle$ are Bloch
wavefunction, {\it i.e.}
$|\alpha_{{\bf k}v}\rangle=e^{i{\bf k\cdot \bf r}}|\tilde\alpha_{{\bf k}v}\rangle/\sqrt{N}$
and $|\beta_{{\bf k}v}\rangle=e^{i{\bf k\cdot \bf r}}|\tilde\beta_{{\bf k}v}\rangle/\sqrt{N}$,
being $N$ the number of unit cells,
$|\tilde\alpha_{{\bf k}v}\rangle$ and $|\tilde\beta_{{\bf k}v}\rangle$
wavefunctions periodic in the lattice, normalized on the unit cell.
In an insulator, the operators
\[
D_{\bf k}=\sum_{v}
|\alpha_{{\bf k}v}\rangle\langle\beta_{{\bf k}v}|
\]
are analytic in ${\bf k}$ and periodic in the reciprocal space.
Cloizeaux has shown in Ref.~\onlinecite{cloizeaux} that an operator having
the properties of $D$ is exponentially localized.

The representation of $\rho^{(n-1)}$ in terms of $D$ is also useful to
obtain a practical expression for the calculation of the
$[{\bf r},\rho^{(n-1)}]$ commutator.
In the limit of a converged ${\bf k}$-points grid,
\[
\frac{1}{N}\sum_{\bf k}
\frac{\partial}{\partial k_\alpha} D_{\bf k}
=\Omega_{\rm c}\int\frac{d^3k}{(2\pi)^3}
~\frac{\partial}{\partial k_\alpha} D_{\bf k}=0,
\]
since the integral over its period of the derivative of a periodic analytic
function is zero.
$\Omega_{\rm c}$ is the unit-cell volume.
From this it can be easily demonstrated that
\begin{equation}
[r_\alpha,D]=\frac{i}{N}
\sum_{{\bf k}v} e^{i{\bf k\cdot \bf r}}
\frac{\partial |\tilde\alpha_{{\bf k}v}\rangle
\langle\tilde\beta_{{\bf k}v}|}{\partial k_\alpha}
e^{-i{\bf k\cdot \bf r}}.
\end{equation}
The terms required in Eq.~(\ref{lin_sys}), when the perturbation is a uniform
electric field, can thus be computed using
\begin{equation}
\langle\psi_{{\bf k}c}^{(0)}|
\left[r_\alpha,D\right]
|\psi_{{\bf k}v}^{(0)}\rangle=
i\sum_{v'}
\langle\tilde\psi_{{\bf k}c}^{(0)}|
\frac{\partial
|\tilde\alpha_{{\bf k}v'}\rangle
\langle\tilde\beta_{{\bf k}v'}|}
{\partial k_\alpha}
|\tilde\psi_{{\bf k}v}^{(0)}\rangle,
\label{pippo}
\end{equation}
where
$|\tilde\psi_{{\bf k}v}^{(0)}\rangle$ are the periodic part of the Bloch
wavefunctions, {\it i.e.}
$|\psi^{(0)}_{{\bf k}v}\rangle=e^{i{\bf k\cdot\bf r}}|\tilde\psi^{(0)}_{{\bf k}v}\rangle/\sqrt{N}$,
and the bra-ket products on the right-hand side are performed on the unit cell.
In practical implementation, the derivative with respect to $k_\alpha$ in
the right-hand side of Eq.~(\ref{pippo}) can be computed numerically by
finite-differentiation, using an expression independent from the arbitrary
wavefunction-phase, as in Refs.~\onlinecite{dalcorso94,nunes01}.

\section{Derivatives of the polarization
\label{sec4}}
Finally, with the present formalism, the computation of the $n$-th order
variation of the polarization density ${\bf P}^{(n)}$, becomes natural.
The components of ${\bf P}^{(n)}$ can be written as:
\begin{equation}
P^{(n)}_\alpha = -\frac{2e}{N\Omega_{\rm c}}Tr\{r_\alpha\rho^{(n)}\},
\label{polariz}
\end{equation}
where the factor two accounts for the spin degeneracy
and $Tr\{A\}$ is the trace of the operator $A$.
We substitute $\rho^{(n)}$ from Eq.~(\ref{rhon}) into
Eq.~(\ref{polariz}).
Using the trace property $Tr\{[A,B]C\}=Tr\{A[B,C]\}$,
is easy to arrive at:
\begin{widetext}
\begin{eqnarray}
P^{(n)}_\alpha &=& -\frac{2e}{N\Omega_{\rm c}}
Tr\left\{
 [r_\alpha,\rho^{(0)}]\rho^{(n)}_{VC}
-[r_\alpha,\rho^{(0)}]\rho^{(n)}_{CV}
+\frac{1}{2} \sum_{i=1}^{n-1}
\left(
[r_\alpha,\rho^{(i)}_{CV}]O^{(n-i)}_{VC}
-[r_\alpha,\rho^{(i)}_{VC}]O^{(n-i)}_{CV}
\right)
\right\}=\nonumber \\
&=&
\frac{2e}{N\Omega_{\rm c}}\sum_{{\bf k}vv'}{\cal I}m
\left(
2\langle\tilde\eta^{(n)}_{{\bf k}v}|
\frac{\partial|\tilde\psi^{(0)}_{{\bf k}v'}\rangle
\langle\tilde\psi^{(0)}_{{\bf k}v'}|
}
{\partial k_\alpha}
|\tilde\psi^{(0)}_{{\bf k}v}\rangle
+\sum_{i=1}^{n-1}
\langle\tilde\chi^{(n-i)}_{{\bf k}v}|
\frac{\partial
|\tilde\eta^{(i)}_{{\bf k}v'}\rangle
\langle\tilde\psi^{(0)}_{{\bf k}v'}|
}
{\partial k_\alpha}
|\tilde\psi^{(0)}_{{\bf k}v}\rangle
\right),
\end{eqnarray}
\end{widetext}
where $n\geq 2$, ${\cal I}m(z)$ is the imaginary part of the complex number
$z$, and we have written the operators $O^{(i)}_{VC}$ as
\[
O^{(i)}_{VC}=\sum_{{\bf k}}
|\psi^{(0)}_{{\bf k}v}\rangle\langle\chi^{(i)}_{{\bf k}v}|,
\]
being
$|\chi^{(i)}_{{\bf k}v}\rangle=O^{(i)}_{CV}|\psi^{(0)}_{{\bf k}v}\rangle$.

\section{Conclusions}
Concluding, we presented a formalism for the calculation of the derivatives of
the electronic density matrix at any order, within the density
functional theory approach.
Beside being simple, this formalism allows the treatment of extended
systems in the presence of an external uniform electric fields in a natural
way, without introducing in the Hamiltonian an additional term depending
on the polarization Berry-phase.

\appendix*
\section{}
The operators defined in Eq.~(\ref{ocv}) are well defined for $\lambda$
sufficiently small, since the series
\[
\frac{1-\sqrt{1-4x}}{2x} = 1 +x +2x^2 +5x^3 +\dots
\]
has the same convergence properties of Eq.~(\ref{taylor}).
From Eq.~(\ref{ocv}) of the text,
$\rho_{CC}=\rho_{CV}O_{VC}=O_{CV}\rho_{VC}$ and
$\rho_{VV}-\rho^{(0)}=-O_{VC}\rho_{CV}=-\rho_{VC}O_{CV}$.
For $n\geq 2$
the $n$-th order variation of $\rho_{CC}$ and $\rho_{VV}$ are:
\[
\rho^{(n)}_{CC}=\sum_{i=1}^{n-1}\rho^{(i)}_{CV}O^{(n-i)}_{VC}=
\sum_{i=1}^{n-1}O^{(n-i)}_{CV}\rho^{(i)}_{VC}
\]
\[
\rho^{(n)}_{VV}=-\sum_{i=1}^{n-1}O^{(n-i)}_{VC}\rho^{(i)}_{CV}=
-\sum_{i=1}^{n-1}\rho^{(i)}_{VC}O^{(n-i)}_{CV}.
\]
Eq.~(\ref{rhon}) of the text easily follows.

Writing $O^{(n)}_{VC}$ as a function of 
$\rho^{(i)}_{CV}$, at the lowest orders we have:
\begin{eqnarray}
O^{(1)}_{VC} &=& \rho^{(1)}_{VC} \nonumber \\
O^{(2)}_{VC} &=& \rho^{(2)}_{VC} \nonumber \\
O^{(3)}_{VC} &=& \rho^{(3)}_{VC} +
\rho^{(1)}_{VC} \rho^{(1)}_{CV} \rho^{(1)}_{VC}
\nonumber \\
O^{(4)}_{VC} &=& \rho^{(4)}_{VC} +
\sum_{i,j,k}^*  \rho^{(i)}_{VC} \rho^{(j)}_{CV} \rho^{(k)}_{VC}
\delta_{i+j+k,4}
+ \nonumber \\
O^{(5)}_{VC} &=& \rho^{(5)}_{VC} +
\sum_{i,j,k}^*  \rho^{(i)}_{VC} \rho^{(j)}_{CV} \rho^{(k)}_{VC}
\delta_{i+j+k,5}
+ \nonumber \\
&&2\rho^{(1)}_{VC} \rho^{(1)}_{CV} \rho^{(1)}_{VC}
 \rho^{(1)}_{CV} \rho^{(1)}_{VC},
\nonumber
\end{eqnarray}
where $\sum^*$ is a sum on positive integers.
These equations allows to compute $\rho^{(n)}$, with $n\leq 6$.


\begin{thebibliography}{99}

\bibitem{DFPT1}
S.\ Baroni, P.\ Giannozzi, and A.\ Testa, 
Phys.\ Rev.\ Lett.\ {\bf 58}, 1861 (1987);
P.\ Giannozzi, S.\ de~Gironcoli, P.\ Pavone, and
S.\ Baroni, Phys.\ Rev.\ B {\bf 43}, 7231 (1991);
S.\ Baroni, S.\ de~Gironcoli, A.\ Dal~Corso, and P.\ Giannozzi,
Rev.\ Mod.\ Phys.\ {\bf 73}, 515 (2001).

\bibitem{DFPT2}
X.\ Gonze and C.\ Lee,  Phys.\ Rev.\ B {\bf 55}, 10355 (1997).

\bibitem{DFT}
P.\ Hohenberg and W.\ Kohn, Phys. Rev. {\bf 136B}, 864 (1964);
W.\ Kohn and L.J.\ Sham, Phys. Rev. {\bf 140A}, 1133 (1965).

\bibitem{NMR}
F.\ Mauri, B.\ Pfrommer, and S.G.\ Louie,
Phys.\ Rev.\ Lett.\ {\bf 77}, 5300 (1996);
C.J.\ Pickard and F.\ Mauri,
Phys.\ Rev.\ B {\bf 63}, 245101 (2001).

\bibitem{gonze95}
X.\ Gonze, Phys.\ Rev.\ A {\bf 52}, 1096 (1995).

\bibitem{levine}
See, for example:
J.E.\ Sipe and Ed.\ Ghahramani, Phys.\ Rev.\ B {\bf 48}, 11705 (1993);
J.L.P.\ Hughes and J.E.\ Sipe, {\it ibid.} {\bf 53}, 10751 (1996);
Z.H.\ Levine and D.C.\ Allan, {\it ibid.} {\bf 44}, 12781 (1991);
Z.H.\ Levine, {\it ibid.} {\bf 49}, 4532 (1994).

\bibitem{dalcorso}
A.\ Dal~Corso and F.\ Mauri,
Phys.\ Rev.\ B {\bf 50}, 5756 (1994);
A.\ Dal~Corso, F.\ Mauri, and A.\ Rubio,
Phys.\ Rev.\ B {\bf 53}, 15638 (1996).

\bibitem{nunes01}
R.W.\ Nunes and X.\ Gonze, Phys.\ Rev.\ B {\bf 63}, 155107 (2001).

\bibitem{kingsmith93}
R.D.\ King-Smith and D.\ Vanderbilt, Phys.\ Rev.\ B {\bf 47}, 1651 (1993).


\bibitem{wannier}
R.W.\ Nunes and D.\ Vanderbilt,
Phys.\ Rev.\ Lett.\ {\bf 73}, 712 (1994).

\bibitem{lazzeri03}
M.\ Lazzeri and F.\ Mauri,
Phys.\ Rev.\ Lett.\ {\bf 90}, 036401 (2003).

\bibitem{pwscf}
S.\ Baroni, A.\ Dal Corso, S.\ de~Gironcoli, and P.\ Giannozzi
(URL http://www.pwscf.org).

\bibitem{abinit}
The ABINIT code is a common project of the Universit\'e Catholique de Louvain,
Corning Incorporated, and other contributors (URL http://www.abinit.org).

\bibitem{nota2}
If one is interested in a perturbation with respect to a macroscopic
uniform electric field, one can set to zero the uniform component of the
Hartree potential in $\partial V^{\rm Hxc}/\partial  E_\alpha$ and in its
higher-order derivatives.~\cite{DFPT1}

\bibitem{dalcorso94}
A.\ Dal~Corso and F.\ Mauri, Phys. Rev. B {\bf 50}, 5756 (1994).

\bibitem{cloizeaux}
J. des~Cloizeaux, Phys. Rev. {\bf 135}, A685 (1964).


\end{thebibliography}
\end{document}